\documentclass[aps,pre,twocolumn,groupedaddress]{revtex4-2}

\usepackage{color}

\usepackage[utf8]{inputenc}
\usepackage[english]{babel}
\usepackage{amsmath,graphicx,enumerate}

\begin{document}

\title{Inhomogeneous entropy production in active crystals with point imperfections} 

\author{L. Caprini$^{1}$}
\author{H. L\"owen$^{1}$}
\author{U. Marini Bettolo Marconi$^{2}$}
\affiliation{$^1$ Heinrich Heine Universit\"at D\"usseldorf, D\"usseldorf, Germany.\\ 
$^2$ Scuola di Scienze e Tecnologie, Universit\`a di Camerino - via Madonna delle Carceri, 62032, Camerino, Italy and
INFN Sezione di Perugia, I-06123 Perugia, Italy}

\date{\today}


\begin{abstract}

The presence of
defects in solids formed by active particles breaks their discrete translational symmetry.
As a consequence, many of their properties become space-dependent and different  from those
characterizing perfectly ordered structures.
Motivated by recent numerical investigations concerning the nonuniform distribution of entropy production
and its relation to the configurational properties of active systems, we 
study theoretically and numerically  
the spatial profile of the entropy production rate when an active solid contains 
an isotopic mass defect.  The theoretical study of such an imperfect active crystal is conducted by
employing a perturbative analysis that considers the perfectly ordered harmonic solid as a reference system.
The perturbation theory predicts a nonuniform profile
of the entropy production extending over large distances from the position of the impurity.
The entropy production rate decays exponentially to its bulk value with a typical healing length that coincides with the correlation length of the spatial velocity correlations characterizing the perfect active solids in the absence of impurities.
The theory is validated against numerical simulations of an active Brownian particle crystal in two dimensions 
with Weeks-Chandler-Andersen repulsive interparticle potential. 
\end{abstract}

\maketitle

\section{Introduction}

Active matter comprehends many systems of biological and technological interest such as bird flocks, cell colonies, spermatozoa, and Janus particles, to mention just a few of them~\cite{marchetti2013hydrodynamics, bechinger2016active}.
All these systems are capable of self-propulsion, namely a mechanism that converts energy from the environment into
directed and persistent motion and drives them out of equilibrium.  Based on experimental evidence, such a self-propulsion or active force is represented at a coarse-grained level by a stochastic process with memory.
In other words, the value of the active force acting on a given particle at a given instant is correlated with the values it took in the past.

In this study, we shall focus on the properties of active matter at high density, a regime characterizing several systems ranging from biological tissues and cell monolayers~\cite{alert2020physical, garcia2015physics, henkes2020dense} populating our skin, to dense colonies of bacteria~\cite{dell2018growing, peruani2012collective,wioland2016ferromagnetic} capable of self-organizing into active two-dimensional crystals of rotating cells~\cite{petroff2015fast}.
Moreover, solid-like configurations have been observed in systems of active Janus colloids~\cite{buttinoni2013dynamical, ginot2018aggregation, palacci2013living, mognetti2013living} and active granular particles~\cite{briand2016crystallization, briand2018spontaneously,baconnier2021selective,plati2019dynamical,plati2022collective}.
Several numerical and theoretical studies investigate the effect of the active force on high-density phases of active matter, such as
liquid, hexatic, and solid~\cite{ferrante2013elasticity, menzel2013traveling, pasupalak2020hexatic, praetorius2018active, caprini2020active, huang2020dynamical, li2021melting,bialke2012crystallization, omar2021phase}.
In two dimensions, the activity shifts the liquid-hexatic and hexatic-solid transition to larger values of the density~\cite{digregorio2018full, klamser2018thermodynamic}, somehow, increasing the effective temperature of the system and broadens the size of the hexatic region that in the passive case is quite narrow~\cite{digregorio2018full}.
More recently, it has been found that dense phases of active matter display spatial velocity correlations~\cite{caprini2020hidden, caprini2020time, marconi2021hydrodynamics, szamel2021long, kuroda2023anomalous, kopp2023spontaneous} a feature absent in equilibrium systems but observed in active glasses~\cite{keta2022disordered, flenner2016nonequilibrium, debets2023glassy}.
This is a phenomenon of dynamical origin determined by the tendency of the particle velocities to align spontaneously even in the absence of direct alignment force~\cite{caprini2020spontaneous}:
it results from the combined action of the persistence of the direction of motion
and the steric repulsion among the particles, while attractive interactions can even induce a flocking transition~\cite{caprini2023flocking}.

To mark the difference between an active and an equilibrium solid with similar structural properties, one can apply the tools of stochastic thermodynamics~\cite{sekimoto2010stochastic,seifert2012stochastic}.
In particular,  the so-called entropy production rate (EPR) provides a quantitative measure of the distance of a system from equilibrium~\cite{fodor2016far, fodor2021irreversibility, o2022time, datta2022second}.
This analysis discriminates between non-equilibrium steady states which produce entropy~\cite{mandal2017entropy, caprini2018comment, pietzonka2017entropy, caprini2019entropy, puglisi2017clausius} and truly equilibrium states whose EPR is zero.
The non-vanishing of the EPR is a universal feature of non-equilibrium systems and occurs when their dynamics break the time-reversal symmetry, i.e. the detailed balance condition is violated.
In the present problem, the system produces both entropy and steady probability currents, a situation that never occurs under equilibrium conditions. 

Being intrinsically out of equilibrium, active matter is an ideal platform to investigate entropy production and shed light on several general properties of non-equilibrium systems.
However, except for specific cases~\cite{cocconi2020entropy, razin2020entropy}, such as non-interacting active particles~\cite{shankar2018hidden, chaki2018entropy} and harmonically confined systems~\cite{garcia2021run, frydel2023entropy}, analytical results for the EPR are difficult to achieve: in general, the EPR for non-linear confining forces or interacting systems has been obtained numerically~\cite{dabelow2021irreversible, grandpre2021entropy}.
Other numerical investigations focused on the study of the EPR of systems spatially inhomogeneous through particle-resolved simulations~\cite{chiarantoni2020work, crosato2019irreversibility, guo2021play} or using field-theoretical descriptions~\cite{nardini2017entropy, borthne2020time, pruessner2022field, paoluzzi2022scaling}.

Recently, we have studied active crystals and found that their vibrational excitations are of two different kinds:
the first is identified with the conventional collective oscillatory modes, known as phonons, and the second describes
additional vibrational excitations, absent at equilibrium and termed entropons because are the modes associated with the entropy production of the system~\cite{caprini2022entropons}.
Under small deviations from equilibrium conditions, entropons coexist without interfering with the conventional phonons, the equilibrium-like excitations. Entropons vanish in equilibrium whereas dominate over phonons when the system is far from equilibrium.
While we have a satisfactory description of the EPR in an ideal solid phase much less is known when its order is altered by the presence of imperfections such as surfaces, defects, or other departures from the perfect periodic arrangement of the active particles.

The aim of this paper is to investigate the EPR in active systems with inhomogeneities
and determine its spatial distribution in the non-equilibrium steady-state. We consider a case that lends itself to analytical and numerical scrutiny: the active crystal containing point imperfections that are due to particles with different masses and destroy the periodic order characterizing perfect crystalline structures. 
This is a classical problem of Solid State Physics~\cite{ziman1972principles} where it is studied to understand the localization
of the phonon modes near the impurity.
We employ this model to shed some light on the EPR in active systems with broken translational invariance
by developing a suitable perturbation theory  around the perfect crystal state by considering a small defect mass. Our method
predicts analytically the spatial profile of the EPR as a function of the distance from the lattice imperfection
 and relates this feature to the existence of a velocity correlation profile.

The paper is structured as follows: in Sec.~\ref{sec:model}, we present the model to describe a solid formed by active particles and calculate the entropy production employing a path-integral technique. Section~\ref{sec:entropyprod} reports the main results of the paper: we derive the perturbative method introduced to calculate the spatial profile of the EPR in the presence of a mass impurity in the solid.
We evaluate explicitly the zeroth-order perturbative EPR, i.e. the entropy production rate of a perfectly ordered crystal, and the first-order 
perturbative correction that describes the effect of the point imperfection that breaks the translational discrete symmetry of the periodic array. Details about the derivations are presented in the appendices to render the exposition.
Finally, the conclusions are presented in Sec.~\ref{sec:conclusions}.

\section{Model}\label{sec:model}
We investigate a solid formed by $N$ ABP's \cite{fily2012athermal, solon2015pressure, siebert2018critical, caprini2020spontaneous, caporusso2020motility, martin2021characterization, vuijk2020lorentz, hecht2022active} in two dimensions, in a square box of size $L\times L$ and apply periodic boundary conditions.
The  evolution of the position, $\mathbf{x}_p$ and velocity, $\mathbf{v}_p=\dot{\mathbf{x}}_p$ of each ABP of mass $m_p$ (with $p=1, ..., N$) 
is governed by the following an underdamped stochastic equation
\begin{equation}
\label{eq:activedynamics}
m_p\dot{\mathbf{v}}_p= -\gamma \mathbf{v}_p + \mathbf{F}_p +\sqrt{2 T \gamma}\, \boldsymbol{\xi}_p + \mathbf{f}^a_p
\end{equation}
where $\boldsymbol{\xi}_p$ is a white noise with zero average and unit variance.
The coefficients $\gamma$ and $T$ are the friction coefficient and the temperature of the solvent bath, respectively.
For equal masses, the ratio, $m\gamma$ corresponds to the typical inertial time,  $\tau_I$, representing the relaxation time of the velocity in equilibrium systems (we remark that in active systems the relaxation of the velocity is determined both by $\tau_I$ and $\tau$~\cite{marconi2021hydrodynamics}).

The active force, $\mathbf{f}^a_p$, provides a certain persistence of the particle trajectory and
drives the system out of equilibrium. In the absence of any other force but the friction, $\mathbf{f}^a_p$ and for  $\tau_I\to 0$
the ABP's would travel at the swim velocity, $v_0$, as shown by the relation:
\begin{equation}
\label{eq:self-propulsiondef}
\mathbf{f}^a_p=\gamma v_0 \mathbf{n}_p \,.
\end{equation}
The stochastic vector $\mathbf{n}=(\cos\theta_p, \sin\theta_p)$ is a unit vector, whose orientation is determined by the angle $\theta_p$, 
subject to Brownian motion
\begin{equation}
\label{eq:dynamica_theta}
\dot{\theta}_p=\sqrt{2D_r} \eta_p \,,
\end{equation}
where $\eta_p$ is a white noise with zero average and unit variance and $D_r$ is the rotational diffusion coefficient.
$D_r$ determines the persistence time of the dynamics $\tau=1/D_r$, i.e. the average time needed by a particle to change direction~\cite{farage2015effective, caprini2022parental}.
 The analysis of the mean square displacement in the independent particle limit leads to the introduction of the so-called active temperature, $T_a= v_0^2 \gamma \tau$, that is an increasing function of both $\tau$ and $v_0^2$.

The force $\mathbf{F}_p$ represents the inter-particle interaction due to a pairwise potential, $U_{\text{tot}}=\sum_{i>j} U(|\mathbf{r}_{i}-\mathbf{r}_{j}|)$, that we choose as a purely repulsive and given by the shift-and-cut WCA potential
\begin{equation}
U(r)=4\epsilon \left[\left(\frac{d_0}{r}\right)^{12}-\left(\frac{d_0}{r}\right)^{6}\right]+\epsilon \,,
\end{equation}
for $r<2^{1/6}$ and zero otherwise.
The parameters $\epsilon$ and $d_0$ represent the energy scale and the particle diameter, respectively.
To consider a solid configuration in numerical simulations, the packing fraction of the system is set to $\phi=\rho d_0^2\pi/4 = 1.1$ that for the range of parameters explored in this study will result in a solid configuration as in the phase diagram reported in Ref.~\cite{caprini2020hidden}.

Above a certain density,
the particles spontaneously arrange themselves to form an almost regular triangular lattice. Assuming that this configuration corresponds to the minimum
of the total potential energy of the system, we Taylor expand $U_{tot}$ around it up to second order in the
displacements, $\mathbf{u}_p$.
 ~\cite{caprini2020hidden, caprini2021spatial}.
 These are defined as the deviations of the particles' coordinates from the perfect lattice positions, $\mathbf{r}^0_p$,  through the relation $\mathbf{u}_p=\mathbf{r}_p-\mathbf{r}^0_p$. Within this approximation,
the force $\mathbf{F}_p$ acting on the $p$ particle reads:
\begin{equation}
\label{eq:forceprofile}
\mathbf{F}_p \approx m\omega^2_E\sum_j (\mathbf{u}_j -\mathbf{u}_p) \,,
\end{equation}
where the sum is restricted to the first neighbor particles and $\omega_E$ is the Einstein frequency of the solid:
\begin{equation}
\omega^2_E=\frac{1}{2m}\left( U''(\bar{x})+\frac{U'(\bar{x})}{\bar{x}}\right) \,.
\end{equation}
$\omega_E$ depends explicitly on the derivatives of $U$ calculated at $\bar{x}$, i.e. the average distance between neighboring particles of the solid (i.e. the lattice constant), which is determined by the packing fraction.

\begin{figure}[!t]
\centering
\includegraphics[width=0.9\linewidth,keepaspectratio] {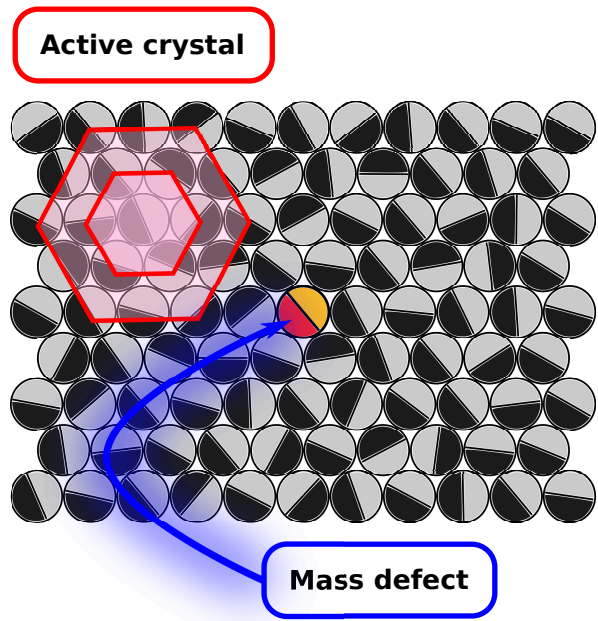}
\caption{A typical solid-like snapshot of the ABP's. The red-yellow particle  has a mass $m+\delta m$,
whereas the remaining grey-black particles have mass $m$. The different colors of each half-disk represent the instantaneous orientation, along the normal to the diameter, of the active force acting on each particle. The hexagons have been drawn to emphasize the
presence of the triangular lattice.
}
\label{fig:Fig0}
\end{figure}

\subsection{Calculation of entropy production}

The stochastic thermodynamics \cite{sekimoto2010stochastic,seifert2012stochastic, speck2016stochastic, szamel2019stochastic} is a powerful tool
to measure how far from thermodynamic equilibrium is a system, i.e. its degree of irreversibility. Such information
is contained in the so-called entropy production rate (EPR), $\dot{s}$, which can be determined by considering the probabilities
of the trajectories connecting two different states of the system.
The EPR 
is expressed in terms of path-probability by resorting to path-integral techniques \cite{spinney2012nonequilibrium, caprini2019entropy, dabelow2019irreversibility,pigolotti2017generic} as
\begin{equation}
\dot{s}=\lim_{t\to\infty}\frac{1}{t} \left\langle\log \left(\frac{\mathcal{P}}{\mathcal{P}_r} \right) \right\rangle \,,
\end{equation}
where the symbol $\langle \cdot \rangle$ represents the steady-state average performed over the realizations of the noise and $\mathcal{P}$ and $\mathcal{P}_r$ are the path probabilities of the forward and backward trajectories of the system, respectively.
These probabilities depend on the whole time history of the dynamical variables of the system $(\mathbf{x}_p, \mathbf{v}_p, \mathbf{f}^a_p)$ conditioned to their initial values $(\mathbf{x}_p(0), \mathbf{v}_p(0), \mathbf{f}^a_p(0))$.
In the case of equilibrium systems, in virtue of the detailed balance condition, which is tantamount to the probabilistic time-reversal symmetry, this ratio is one and the EPR vanishes.
To estimate $\mathcal{P}$ and $\mathcal{P}_r$, let us remark that the probability of the trajectory of a stochastic system is uniquely determined by the probability of observing a path-trajectory of the stochastic noises, that in our case have a Gaussian distribution.
Therefore, one performs a transformation from the noise variables to the dynamical variables by using the equation of motion~\eqref{eq:activedynamics} together with Eq.~\eqref{eq:dynamica_theta}. In doing so, we neglect the determinant of the transformation because, in the present case of additive noise, this term does not contribute to the EPR.
Applying this procedure, the probability of forward and backward trajectories are expressed as $\mathcal{P}\sim e^{\mathcal{A}}$ and $\mathcal{P}_r\sim e^{\mathcal{A}_r}$, respectively, where $\mathcal{A}$ and $\mathcal{A}_r$ are actions associated to backward and reverse dynamics.
The action $\mathcal{A}$ is obtained by expressing the Gaussian distribution of the noise variables, $ \boldsymbol{\xi}_p$, in terms
of the state variables $(\mathbf{x}_p, \mathbf{v}_p, \mathbf{f}^a_p)$
using  the relation between the two sets of variables given by Eq. \eqref{eq:activedynamics} with the result:
\begin{equation}
\mathcal{A}=-\sum_p\frac{m_p^2}{4T\gamma}\int dt \left[ \dot{\mathbf{v}}_p - \frac{\mathbf{F}_p}{m_p} - \frac{\mathbf{f}^a_p}{m_p} + \frac{\gamma}{m_p}\mathbf{v} _p\right]^2 
\end{equation}
Here, we have not included in the action the contribution associated with the rotational noise,  $\eta_p$, since it is known that the 
simple Brownian process of Eq.~\eqref{eq:dynamica_theta} does not generate entropy.
The action of the backward trajectory $\mathcal{A}_r$ can be obtained by applying the time-reversal transformation to the dynamics~\eqref{eq:activedynamics} and considering the parity of the dynamical variables, $(\mathbf{x}_p, \mathbf{v}_p, \mathbf{f}^a_p)$, under this transformation.
By denoting with the subscript $r$ the time-reversed variables, we assume:
\begin{subequations}
\begin{align}
&\mathbf{x}_r \to \mathbf{x}\\
&\mathbf{v}_r \to -\mathbf{v}\\
&\mathbf{f}^a_r \to \mathbf{f}^a
\end{align}
\end{subequations}
for each particle (above, the particle index has been suppressed for notational convenience).
In this way, the backward action, $\mathcal{A}_r$, reads
\begin{equation}
\mathcal{A}_r=-\sum_p\frac{m_p^2}{4T\gamma}\int dt \left[ \dot{\mathbf{v}}_p - \frac{\mathbf{F}_p}{m_p} - \frac{\mathbf{f}^a_p}{m_p} - \frac{\gamma}{m_p}\mathbf{v} \right]^2 \,,
\end{equation}
where we have again neglected the irrelevant contribution of the angular dynamics, for the same reasons given above.

Performing algebraic calculations, the expression for $\dot{s}$ can be analytically derived and reads:
\begin{equation}
\label{eq:exact_general_entropyprod}
\dot{s} = \sum_p \dot{s}_p
\end{equation}
where $\dot{s}_p$ is the entropy production generated by a single particle given by
\begin{equation}
\dot{s}_p = \frac{1}{T}\langle \mathbf{f}^a_p \cdot \mathbf{v}_p\rangle + {\text{b.t.}} \,.
\end{equation} 
The expression ${\text{b.t.}}$ means boundary terms, i.e. the additional terms that do not contribute to the steady-state entropy production and vanish in the long-time limit.
Such a result holds, in general, for underdamped active particles subject to a persistent active force and in contact with a thermal bath, independently of their density and of the dimension of the system.

By introducing the spatial Fourier transform of the dynamical variables denoted by hat-symbols, and by neglecting boundary terms, $\dot{s}_p$ can be calculated in the Fourier space, in particular, in the  frequency domain, obtaining
\begin{equation}
\label{eq:entropyprod}
\begin{aligned}
\dot{s}_p &= \lim_{t\to\infty}\frac{1}{t}\int \frac{d\omega}{2\pi} \frac{\langle \hat{\mathbf{f}}^a_p(-\omega) \cdot \hat{\mathbf{v}}_p(\omega)\rangle}{T} + c.c.\\
\end{aligned}
\end{equation}
where $\hat{\mathbf{f}}^a_p(\omega)$ and $\hat{\mathbf{v}}_p(\omega)$ are the Fourier transforms in the frequency domain of the active force and velocity of the $p$ particle, respectively, defined in appendix~\ref{appendix:spatial_fouriertransform} and the symbol "$c.c$"
stands for complex conjugate.
The EPR, $\dot{s}$, is proportional to the frequency integral of the real part of the cross-correlation between the Fourier components of active force and velocity.
We also introduce the spectral entropy $\sigma_{p}(\omega)$ for each particle as the integrand in Eq.~\eqref{eq:entropyprod}
\begin{equation}
\sigma_{p}(\omega) = \lim_{t\to\infty} \frac{1}{t} \frac{\langle \hat{\mathbf{v}}_p(\omega)\cdot \hat{\mathbf{f}}^a_p(-\omega)\rangle}{T} + c.c.
\end{equation}
Note that this term corresponds to the spectral dissipation of the particle $p$ due to the active force, divided by the temperature of the bath.


\section{Entropy production of active solids with impurities}\label{sec:entropyprod}

A real solid may contain various kinds of imperfections or surfaces which affect the properties of the perfect crystal. For the sake of simplicity, we confine ourselves to isolated defects such as substitutional particles of different mass \cite{ziman1972principles}.

To understand the effect of substitutional impurities on the EPR, we modify the mass of one particle, setting $m_1 \neq m$ and $m_{p}=m$ for all remaining particles. 
Since this operation breaks the discrete translational symmetry of the lattice, both the displacement and the EPR, $\dot{s}_p$, 
become position dependent.
In the continuum limit,  the local EPR, $\dot{s}_p \to \dot{s}(r)$, becomes a function of the distance $r$ from the location of the impurity.
A sketch of the model is shown in Fig.~\ref{fig:Fig0}.
Notwithstanding the problem described by Eqs.\eqref{eq:activedynamics} and \eqref{eq:forceprofile} is linear and one could use a numerical matrix inversion method to determine with great accuracy the solution  $\mathbf{u}_p$ and $\dot{s}(r)$, 
we are interested in
 getting explicit  predictions. Therefore, we apply an  analytical perturbative approach choosing the mass difference $\delta m=(m_1-m)$ as a small parameter.

\subsection{General strategy of the perturbation scheme}

To proceed analytically, we choose $m_1= m+\delta m$ with $|\delta m|\ll m$ and apply a perturbative method by expanding the solution in powers of the small parameter $\delta m/m \ll 1$.
Explicitly the modified equation of motion for the imperfect lattice reads
\begin{equation}
\label{eq:impurityequation}
(m+\delta_{p0} \delta m)\dot{\mathbf{v}}_p= -\gamma \mathbf{v}_p + \mathbf{F}_p +\sqrt{2 T \gamma} \boldsymbol{\xi}_p + \mathbf{f}^a_p
\end{equation}
where the Kronecker delta function $\delta_{p0}$ selects the particle $p=0$ corresponding to the imperfection. The force $\mathbf{F}_p$ is approximated within the harmonic approximation of Eq.~\eqref{eq:forceprofile}.

By introducing the continuous Fourier transforms in the frequency domain $\omega$ of the displacement
$\hat{\mathbf{u}}_p=\hat{\mathbf{u}}_p(\omega)$,  the velocity $\hat{\mathbf{v}}_p=i\omega\hat{\mathbf{u}}_p(\omega)$, the
active force $ \hat{\mathbf{f}}^a_p(\omega)$ and the white noise $\hat{\boldsymbol{\xi}}_p(\omega)$
Eq. \eqref{eq:impurityequation} can be rewritten as
\begin{equation}
\label{eq:Fourierspace_dynamics}
L_{pk}(\omega) \hat{\mathbf{u}}_k - \delta m \omega^2 \hat{\mathbf{u}}_0 \delta_{p0} = \hat{\mathbf{f}}^a_p + \sqrt{2 \gamma T} \hat{\boldsymbol{\xi}}_p
\end{equation}
where the time-Fourier transform of a generic observable is denoted by the hat symbol and
the Einstein  summation convention used.
The matrix  elements $L_{pk}(\omega)$ have the following form
\begin{equation}
L_{pk}(\omega) = (-m\omega^2 + i \omega \gamma) \delta_{pk} - m\omega^2_E \sum_j^*\delta_{p,k+j}
\end{equation}
where the sum over the index $j$ runs over the nearest neighbors $(k+j)$  of the particle $p$. When $\delta m=0$,   Eq.~\eqref{eq:Fourierspace_dynamics} corresponds to the dynamics of a perfect lattice and can be 
rewritten with the help  of the lattice Green's function, $G_{pn}(\omega)$, in the $\omega$-representation
as:
\begin{equation}
\label{eq:Fourierspace_dynamicsb}
\hat{\mathbf{u}}_p(\omega)=G_{pn}(\omega) ( \hat{\mathbf{f}}^a_n + \sqrt{2 \gamma T} \hat{\boldsymbol{\xi}}_n)\, .
\end{equation}
where $G_{pn}(\omega)$ is expressed with the help of
the discrete spatial-Fourier transform:
\begin{equation}
G_{pn}(\omega)= L^{-1}_{pn}(\omega) = \frac{1}{N} \sum_{\mathbf{q}} \frac{e^{-i \mathbf{q}\cdot (\mathbf{r}^0_{\mathbf{n}}-\mathbf{r}^0_{\mathbf{p}}})  }{-m \omega^2 + i\omega \gamma + m \omega^2(\mathbf{q})} \,,
\label{eq:greenlattice}
\end{equation}
where the sum runs over the dimensionless reciprocal lattice wave vectors $\mathbf{q}$
(see appendix \ref{appendix_pntegrals_omega}). The quantity
$\omega(\mathbf{q})$ represents the dispersion relation of the vibrational modes  of the lattice and depends on the Einstein frequency, $ \omega_E$  and on the lattice structure. It is obtained by solving the secular equation associated with the lattice 
harmonic oscillations~\cite{ashcroft2022solid}.
Since the perturbative method discussed below is independent of the specific lattice structure, being
the relative information contained in $\omega(\mathbf{q})$,
we postpone (see Eq.~\eqref{eq:omegaq_twoDtriangular}) the presentation of the explicit expression of $\omega(\mathbf{q})$ 
for the case of the triangular lattice in two dimensions and nearest neighbor interactions.
By solving Eq.~\eqref{eq:Fourierspace_dynamics} when $\delta m\neq 0$, we obtain
\begin{equation}
\label{eq:dynamics_omega_pert}
\hat{\mathbf{u}}_p(\omega)= G_{pn}(\hat {\mathbf{f}}^a_n + \sqrt{2\gamma T} \hat{\boldsymbol{\xi}}_{n})+ \delta m \omega^2 G_{p0}\,\hat{\mathbf{u}}_0 \,.
\end{equation}
The entropy production $\dot{s}$ can now be calculated by using Eq.~\eqref{eq:entropyprod}, which requires the knowledge of the cross-dynamical correlation between the active force and velocity.
We multiply Eq.~\eqref{eq:dynamics_omega_pert} by $\mathbf{f}^a_n(-\omega)$, take the average over the realizations of the noise, and obtain
\begin{equation}
\label{eq:eq_corr_dynamic}
\begin{aligned}
\langle \hat{\mathbf{u}}_p(\omega) \hat{\mathbf{f}}_p^a(-\omega)\rangle =& G_{pn}(\omega) \langle \hat{\mathbf{f}}_n^a(\omega) \hat{\mathbf{f}}_p^a(-\omega) \rangle \\
&+\delta m \,\omega^2 G_{pn}\langle \hat{\mathbf{u}}_n(\omega) \hat{\mathbf{f}}_p^a(-\omega) \rangle \delta_{n0} \,.
\end{aligned}
\end{equation}
Multiplying both sides 
of Eq.~\eqref{eq:eq_corr_dynamic} by $(i\omega)$ 
we find a relation for the average $\langle \hat{\mathbf{v}}_p(\omega) \hat{\mathbf{f}}_p^a(-\omega)\rangle$.
To proceed further, we need the explicit expression of the correlation $\langle \hat{\mathbf{f}}^a_p(\omega) \hat{\mathbf{f}}_p^a(-\omega)\rangle$, which is estimated by approximating the ABP dynamics with the AOUP model~\cite{szamel2014self, maggi2015multidimensional, szamel2015glassy, wittmann2017effective, caprini2018activeescape}.
This strategy has been often adopted with success in the literature to get analytical predictions~\cite{das2018confined, caprini2018active, fodor2016far, maggi2020universality}.
We obtain
\begin{equation}
\label{eq:fafa_corr}
\lim_{t\to\infty}\frac{1}{t} \langle \hat{\mathbf{f}}^a_n(\omega) \hat{\mathbf{f}}_p^a(-\omega)\rangle = 2 v_0^2 \gamma^2 \frac{\tau}{1+\omega^2\tau^2} \delta_{np}\,,
\end{equation}
and
\begin{equation}
\label{eq:fafa_average}
 \langle \hat{\mathbf{f}}^a_p(\omega) \rangle=0
\end{equation}
as discussed in Appendix~\ref{appendix:AOUP}.
By replacing the expression~\eqref{eq:fafa_corr} in Eq.~\eqref{eq:eq_corr_dynamic}, we finally get the equation describing the dynamical cross correlation between active force and velocity
\begin{equation}
\begin{aligned}
\label{eq:corr_for_EP}
\lim_{t\to\infty} \frac{1}{t}&\langle \hat{\mathbf{v}}_p(\omega) \hat{\mathbf{f}}_p^a(-\omega)\rangle =2 i \omega G_{np}(\omega)  v_0^2 \gamma^2 \frac{\tau}{1+\omega^2\tau^2} \delta_{np} \\
&+ \lim_{t\to\infty} \frac{i}{t}\delta m\, \omega^3 G_{p0}(\omega)\langle \hat{\mathbf{u}}_0(\omega) \hat{\mathbf{f}}_p^a(-\omega) \rangle \,.
\end{aligned}
\end{equation}
By dividing by the temperature $T$ and taking the real part of Eq.~\eqref{eq:corr_for_EP}, we obtain the equation to determine $\sigma_p(\omega)$, that reads
\begin{equation}
\label{eq:sigma_notclosedresult}
\begin{aligned}
T&\sigma_{p}(\omega) = - 2\omega  v_0^2 \gamma^2 \frac{\tau}{1+\omega^2\tau^2} \text{Im}[G_{np}(\omega)] \delta_{np}\\
&- \lim_{t\to\infty} \frac{1}{t}\delta m  \omega^3 \text{Im} [G_{p0}(\omega)\langle \hat{\mathbf{u}}_0(\omega) \hat{\mathbf{f}}_p^a(-\omega) \rangle ] \,.
\end{aligned}
\end{equation}
Equation~\eqref{eq:sigma_notclosedresult} is not closed and cannot be used to determine the entropy production rate because contains the dependence on the unknown correlation $\langle \hat{\mathbf{u}}_0(\omega) \hat{\mathbf{f}}_p^a(-\omega) \rangle$.
To proceed we employ a perturbative method and expand the solution in powers of $\lambda=\delta m/m \ll 1$:
\begin{equation}
\hat{\mathbf{u}}_p(\omega) = \hat{\mathbf{u}}_p^{(0)}(\omega) + \lambda\hat{\mathbf{u}}_p^{(1)}(\omega) + \lambda^2 \hat{\mathbf{u}}_p^{(2)}(\omega) + ... \,.
\end{equation}
where the superscript $^{(n)}$ denotes the order of the perturbative solution that is consistent with a perturbative solution of the spectral entropy production:
\begin{equation}
{\sigma}_p(\omega) = {\sigma}_p^{(0)}(\omega) + \lambda {\sigma}_p^{(1)}(\omega) + \lambda^2 {\sigma}_p^{(2)}(\omega) + ... \,.
\end{equation}
and of its integral over $\omega$, i.e. $\dot{s}$, so that
\begin{equation}
\dot{s}_p = \dot{s}_p^{(0)} + \lambda \dot{s}_p^{(1)} + \lambda^2 \dot{s}_p^{(2)} + ... \,.
\end{equation}
Note that the zeroth-order entropy production is independent of the particle index $p$, which can be dropped at this order.
Instead, the EPR is spatially dependent starting from the first correction in $\delta m/m$.

In our discrete formalism, this implies that $\dot{s}_p^{(0)}=\dot{s}^{(0)}$ and $\sigma_p^{(0)}(\omega)=\sigma^{(0)}(\omega)$ for every $p$, while, in general, we expect a dependence on $p$, or in other words a spatial dependence, only starting from the first correction in $\delta m/m$.

\subsection{Zeroth-order result: the homogeneous solid}\label{subsec:homogeneous}

The zero-order solution of Eq.~\eqref{eq:sigma_notclosedresult} (obtained for $\delta m=0$) corresponds to the spectral entropy, ${\sigma}^{(0)}(\omega)$, of the homogeneous solid, in the absence of the impurity.

Setting $\delta m=0$ in Eq.~\eqref{eq:sigma_notclosedresult}, we obtain the zeroth-order value of the spectral entropy production per particle
\begin{equation}
\begin{aligned}
T\sigma^{(0)}(\omega) = & - 2 \omega v_0^2 \gamma^2 \frac{\tau}{1+\omega^2\tau^2} \text{Im}[G_{00}(\omega)]  \,,
\end{aligned}
\end{equation}
which is independent of the position.
By integrating over $\omega$, we get the zeroth-order expression for the entropy production rate 
\begin{equation}
\begin{aligned}
\label{eq:i2integral2}
T\dot{s}^{(0)} &= \frac{2}{N} \sum_{q} \int_{-\infty}^\infty \frac{d\omega}{2\pi} \frac{\tau\gamma}{1+\omega^2\tau^2} \frac{\omega^2 v_0^2 \gamma^2}{m^2 (\omega^2(\mathbf{q})- \omega^2)^2 + \omega^2 \gamma^2 }\\
&=\frac{v_0^2 \tau \gamma}{\tau + \tau_I} \mathcal{G}_{00}.
\end{aligned}
\end{equation}
 The second equality introduces the propagator, $\mathcal{G}_{00}$, whose expression is:
\begin{equation}
\mathcal{G}_{00}= \frac{1}{N}\sum_q \frac{1}{1+ \frac{\tau^2\tau_I}{\tau+\tau_I} \omega^2(\mathbf{q})}\,.
\label{eq:g00}
\end{equation}
The $\omega$-integration in Eq.\eqref{eq:i2integral} leading to Eq.\eqref{eq:g00}  is reported in appendix 
\ref{Integralsfrequency} (see Eq.\eqref{eq:i2integral}). In practice, we convert the sum over wave-vectors into an integral over the Brillouin zone
of volume $\Omega$ by replacing $ \frac{1}{N}\sum_q \to  \int_{\Omega} \frac{d{\mathbf{q}}}{\Omega} $ and
 $\mathcal{G}_{00}\to\mathcal{G}(0)$.
 The prefactor in Eq.\eqref{eq:i2integral2} is identified with the EPR of the non-interacting system, $\dot{s}_{\text{free}}$, 
 according to the relation:
\begin{equation}
\dot{s}_{\text{free}}= \frac{v_0^2 \tau\gamma}{T}\frac{1}{\tau_I+\tau} \,,
\end{equation}
and rewrite:
\begin{equation}
\label{eq:dots_zeroorder_corr}
\dot{s}^{(0)}= \dot{s}_{\text{free}}\mathcal{G}(0)\, .
\end{equation}
The quantity $ \dot{s}_{\text{free}}$
is a function of the ratio between the active temperature, $v_0^2 \gamma \tau$, and the thermal temperature. At fixed active temperature, it decreases as the persistence time, $\tau$, and inertial time, $\tau_I$ increase .
Instead, the term $\mathcal{G}_{00}$ accounts for the interactions among the particles in the solid, is $1$
in the non-interacting limit, and $\mathcal{G}_{00}\leq1$ for the interacting case since
$\omega^2(\mathbf{q}) \geq 0$.
In other words, the interaction decreases the value of the EPR with respect to the free-particles case
because the interactions characterizing the solid hinder the particle's ability to move with the same speed as free particles so that $|\mathbf{v}| \ll v_0$.
As a consequence, the entropy production of a solid formed by $N$ active particles is always smaller than the entropy production of $N$ potential-free active particles, $\dot{s} \leq \dot{s}_{\text{free}}$.

\subsubsection{Explicit evaluation of the zeroth-order correction}

To obtain an explicit expression for the zeroth order EPR of Eq.\eqref{eq:i2integral2}
we, now, compute analytically $\mathcal{G}(0)$ whose value depends on the dimension of the system and the lattice structure. 
In the case of a triangular lattice, which is the structure found in our simulations at high-density, the dispersion relation
reads:
\begin{equation}
\label{eq:omegaq_twoDtriangular}
\omega^2(\mathbf{q})=2\omega_E^2\left(3 -\cos(q_x\bar{x}) - 2 \cos\left( q_x \frac{\bar{x}}{2} \right)\cos\left( \frac{\sqrt{3}}{2}q_y \bar{x} \right) \right).
\end{equation}
The summand appearing in Eq.\eqref{eq:g00} when $|\mathbf{q}|\to 0$ becomes of the Ornstein-Zernike form
$\propto (1+\xi^2 q^2)^{-1}$ and thus we can define a correlation length
 $\xi$ from the relation:
\begin{equation}
\label{eq:correlationlength_2D}
\xi^2= \frac{3}{2} \bar{x}^2 \frac{\tau^2}{1+\tau/\tau_I }\omega_E^2\, .
\end{equation}
This length coincides with the correlation length of the spatial velocity correlation of an active solid, $\langle \mathbf{v}(\mathbf{r}) \cdot\mathbf{v}(0) \rangle$ and, as already discussed in Ref.\cite{caprini2020active}, is an increasing function of $\tau$ and of $\tau_I$.
The integral can be computed exactly as shown in the appendix  \ref{appendix:change}  where we find
\begin{equation}
\label{eq:G0principal}
\mathcal{G}(0)=\frac{1}{1+\xi^2} \frac{6}{\pi z \sqrt{c}} K[k]
\end{equation}
where the parameters $c,z$ and $k$ are also given in appendix and $K[k]$ is the complete elliptic integral of the first kind
\cite{abramowitz1988handbook}.

\begin{figure}[!t]
\centering
\includegraphics[width=0.9\linewidth,keepaspectratio]{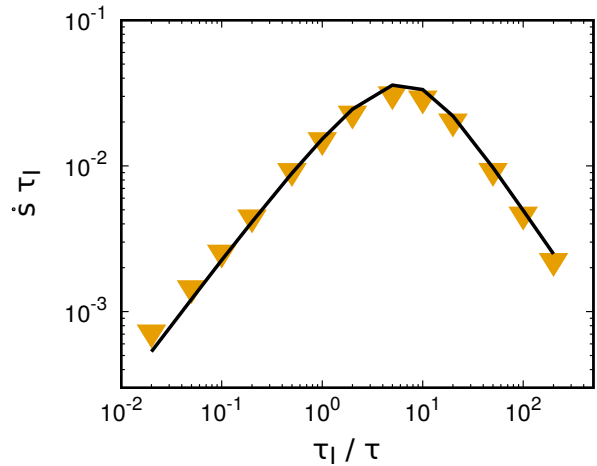}
\caption{
EPR of a perfect  two-dimensional crystal with triangular structure, $\dot{s}$, rescaled by the inertial time, $\tau_I=m/\gamma$, as a function of the reduced inertial time $\tau_I/ \tau)$.
Points are obtained by numerical simulations obtained by integrating the dynamics~\eqref{eq:activedynamics} in the absence of defects, i.e. $m_p=m$ for every $i$, while the solid black line is calculated from the theoretical prediction~\eqref{eq:dots_zeroorder_corr}.
The parameters of the simulations are $N=10^4$ and $\phi=1.1$.
}
\label{fig:Fig1}
\end{figure}

In Fig.~\ref{fig:Fig1}, the theoretical EPR, $\dot{s}^{(0)}$, is compared with the one obtained in numerical simulations of
the solid phase. Results are plotted as a function of the rescaled inertial time $\tau_I/\tau$: $\dot{s}^{(0)}$
increases from zero, attains a maximum value before vanishing for large values of $\tau_I/\tau$,
a behavior consistent with the Clausius inequality, $\dot{s}\geq0$, being $\dot{s}=0$  at equilibrium.
In fact, when the persistence time, $\tau$, is the shortest time scale of the system, 
($\tau_I/\tau\to \infty$), the active force $\mathbf{f}^a_p$ can be assimilated to a Brownian process (persistence time $\approx 0$)  whose EPR is null: it is easy to verify from Eqs.\eqref{eq:g00} and
\eqref{eq:dots_zeroorder_corr} that when  $\xi \to 0$ also $\dot{s}\to 0$ because the factor $\dot{s}_{\text{free}}$ vanishes.
This situation corresponds to an underdamped colloidal solid under equilibrium conditions, described by standard Boltzmann statistics. 
 By decreasing $\tau_I/\tau$ (i.e. increasing the persistence time), the system departs from equilibrium and $\dot{s}$ increases.
For small deviations from equilibrium, the growth of $\dot{s}$ is essentially determined by the factor $\dot{s}_{\text{free}}$ because $\mathcal{G}(0)$ remains close to $1$ and
the EPR scales as $\dot{s}\approx \dot{s}_{\text{free}} \sim \tau/(\tau+\tau_I) \approx \tau/\tau_I$ for $\tau_I /\tau \gg 1$.
Such a linear increase continues up to values of $\tau_I /\tau$ where  $\mathcal{G}(0)$
sensibly departs from $1$.
This situation occurs when the size of coherent domains of the velocities becomes relevant, as revealed by the increasing velocity correlations.
Indeed, when the correlation length of these domains reaches the size of the particle diameter, $\xi \approx \sigma$, the entropy production rate attains its maximum.
A further decrease of $\tau_I/ \tau$ leads to the decrease of $\dot{s}$
because the most relevant contribution to the EPR stems from the boundaries of the domains, 
whereas the particles (whose velocities are more aligned)  located in their interior
 provide less relevant contributions.
As a consequence, $\dot{s}\to0$ when $\tau_I/ \tau \to 0$, 
the limit where $\xi\to\infty$.
We remark that this decrease is in agreement with the presence of arrested states observed numerically in systems of dense active particles in the infinite persistence time limit~\cite{casiulis2021self, debets2021cage} for which $\dot{s}\approx0$.

\subsection{First-order result: the effect of an impurity}

Imperfections always break the discrete spatial translational symmetry of the crystal.
Hence, some observables, including the local entropy production rate, are expected to become spatially dependent on the distance from the defect and take a constant value  away from it,  the one characterizing the perfect crystal.

To analytically predict the local EPR, we employ the  first-order perturbative expression, $\sigma_p^{(1)}(\omega)$, given by Eq.~\eqref{eq:sigma_notclosedresult}:
\begin{equation}
\label{eq:first_orderEP}
\begin{aligned}
T\sigma_p^{(1)}(\omega) = &- m  \omega^3 \text{Im} [G_{p0}(\omega)\langle \hat{\mathbf{u}}^{(0)}_0(\omega) \hat{\mathbf{f}}_p^a(-\omega) \rangle ] \,,
\end{aligned}
\end{equation}
where  $\hat{\mathbf{u}}^{(0)}_0(\omega)$ is the unperturbed displacement at the location of the impurity.
 Eq.~\eqref{eq:dynamics_omega_pert} gives the expression
$\hat{\mathbf{u}}^{(0)}_0(\omega)= G_{0n}(\hat {\mathbf{f}}^a_n + \sqrt{2\gamma T} \hat{\boldsymbol{\xi}}_{n})$
that substituted in Eq.~\eqref{eq:first_orderEP} yields:
\begin{equation}
\label{eq:first_orderEP_2}
\begin{aligned}
T\sigma_p^{(1)}(\omega) = &- m  \omega^3 \text{Im} [G_{p0}(\omega) G_{0n}(\omega)  \lim_{t\to\infty} \frac{1}{t} \langle \hat{\mathbf{f}}^a_n(\omega) \hat{\mathbf{f}}_p^a(-\omega) \rangle ]\\
=&- m  \omega^3 \text{Im} [G_{p0}(\omega) G_{0p}(\omega)] \frac{2 v_0^2 \gamma^2 \tau}{1+\omega^2\tau^2} \, .
\end{aligned}
\end{equation}
where we employed the correlator of the active force, Eqs.\eqref{eq:fafa_corr}, to obtain the last equality.

After replacing the expression for $G_{0p}(\omega)$ and $G_{p0}(\omega)$ using Eq.\eqref{eq:greenlattice}
 and integrating over the frequency $\omega$, we obtain the first-order perturbative correction to the EPR.
Since the calculations are rather lengthy, here, we show only the final results while details of the derivation are reported in Appendix~\ref{appendix_pntegrals_omega}:
\begin{equation}
\label{eq:first_order_correction_EP}
\dot{s}^{(1)}_p = - \frac{v_0^2 \tau \gamma \tau_I}{T(\tau + \tau_I)^2}
\mathcal{G}_{0p} \mathcal{G}_{p0} = - \frac{ \dot{s}_{\text{free}}\tau_I}{\tau + \tau_I}
\mathcal{G}_{0p} \mathcal{G}_{p0} \,,
\end{equation}
where
\begin{equation}
\label{eq:I0p}
\mathcal{G}_{0p} =\frac{1}{N} \sum_q \frac{e^{-i \mathbf{q}\cdot \mathbf{r}_p^0}}{1+ \frac{\tau^2\tau_I}{\tau+\tau_I} \omega^2(\mathbf{q})}
\end{equation}
and $\mathcal{G}_{p0}=\mathcal{G}^*_{0p}$.
By approximating the sum $\sum_{q}$ by an integral, 
\begin{equation}
\label{eq:I0p2}
\begin{aligned}
\mathcal{G}_{0p}\to \mathcal{G}(r)=&\int_{\Omega} \frac{d{\mathbf{q}}}{\Omega} \frac{ e^{-i \mathbf{q}\cdot \mathbf{r}_p^0}}{1+ \frac{\tau^2\tau_I}{\tau+\tau_I} \omega^2(\mathbf{q})} \,.
\end{aligned}
\end{equation}
The above integral depends on the distance from the impurity, $r$ and  converges to the value $\mathcal{G}(0)$ as $r\to 0$.
Thus, the first-order correction to the entropy production, at distance $r$ from the impurity, becomes:
\begin{equation}
\dot{s}^{(1)}_p \to \dot{s}^{(1)}(r) = -\dot{s}_{\text{free}} \mathcal{G}(r)^2 \frac{\tau_I}{\tau_I +\tau}\,.
\label{eq:first_order_correction_EP2} 
\end{equation}

.

\begin{figure}[!t]
\centering
\includegraphics[width=0.9\linewidth,keepaspectratio]{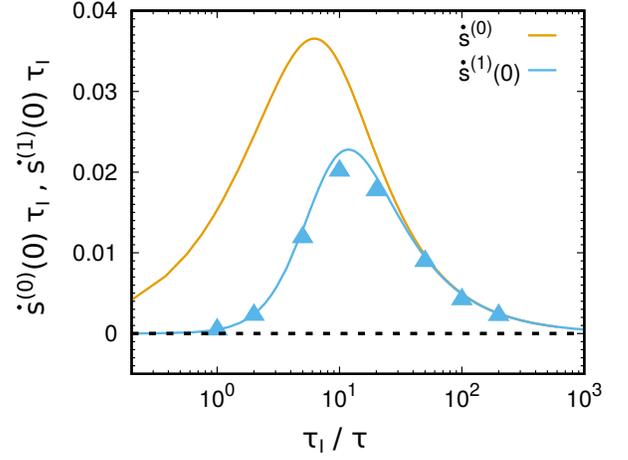}
\caption{
Entropy production rate, $\dot{s}(0)$ (rescaled by $\tau_I=m/\gamma$) calculated at the position of the defect, as a function of the reduced inertial time, $\tau_I/\tau$.
The yellow curve denotes the zeroth-order prediction $\dot{s}^{(0)}$ i.e. the bulk value of the EPR theoretically predicted (see Eq.~\eqref{eq:dots_zeroorder_corr}).
The blue light curve and the light blue points are the first-order correction of the entropy production, $\dot{s}^{(1)}(0)$, calculated at the same position. The solid line represents the theoretical predicition~\eqref{eq:prediction_s1peak}, while points are obtained from numerical simulations conducted by integrating the dynamics~\eqref{eq:activedynamics} with $|\delta m|/m=0.2$.
The values of the blue points,
$\dot{s}^{(1)}(0)$, are estimated from the difference $\dot{s}(0)-\dot{s}^{(0)}$ between the theoretical bulk value
$\dot{s}^{(0)}$  and the numerical value of $\dot{s}(0)$.
The parameters of the simulations were fixed at $N=10^4$ and $\phi=1.1$.
}\label{fig:Fig2}
\end{figure}

According to Eq.~\eqref{eq:first_order_correction_EP2} 
the first order perturbative correction to the EPR at  the position of the impurity, $r=0$, is
\begin{equation}
\label{eq:prediction_s1peak}
\dot{s}^{(1)}(0) = - \frac{ \dot{s}_{\text{free}}\tau_I}{\tau + \tau_I}
[\mathcal{G}(0)]^2 = - \dot{s}^{(0)} \mathcal{G}(0)\frac{\tau_I}{\tau + \tau_I}\,.
\end{equation}
We  estimated numerically
$\dot{s}^{(1)}(0)$ from the difference $(\dot{s}(0) - \dot{s}^{(0)})$, i.e. by subtracting the zeroth-order EPR from the total EPR obtained from the simulations. The resulting value of
$\dot{s}^{(1)}(0)$ together with the zeroth-order theoretical prediction $\dot{s}^{(0)}$, shown for comparison,
is plotted in Fig.~\ref{fig:Fig2} as a function of the reduced inertial time $\tau_I/\tau$.
Both $\dot{s}^{(0)}$ and  $\dot{s}^{(1)}(0)$  are bell-shaped curves
but the two maxima do not coincide.
Indeed, $\dot{s}^{(1)}(0)$ tends to zero as equilibrium is approached, i.e. when $\tau_I/\tau \gg 1$.
By decreasing the ratio $\tau_I/\tau$ the system departs from equilibrium and $\dot{s}^{(1)}(0)$ grows, as $\dot{s}^{(0)}$ also does, and reaches a peak for $\tau_I/\tau \approx 10$. Such an increase when $\tau_I /\tau$ decreases is mainly due to the prefactor $\dot{s}_{\text{free}}$ in Eq.~\eqref{eq:prediction_s1peak}.

In the large persistence regime, $\tau_I/\tau \leq 10$, the first-order correction $\dot{s}^{(1)}(0)$ displays a decrease similar to $\dot{s}^{(0)}$.
From Eq.~\eqref{eq:prediction_s1peak} we see that
$
\left|\frac{\dot{s}^{(1)}(0)}{\dot{s}^{(0)}}\right| = \mathcal{G}(0)\frac{\tau_I}{\tau + \tau_I} \leq 1\,$.
The inequality holds because both $\mathcal{G}(0)$ and $\tau_I/(\tau+\tau_I)$ are less or equal to 1.
Moreover, $\frac{\dot{s}^{(1)}(0)}{\dot{s}^{(0)}}$ is a decreasing function of $\tau$ and an increasing function of $\tau_I$.
In the limit of infinite $\tau$, the system approaches an arrested state (with decreasing speed, $|\mathbf{v}|$), and, as a consequence, not only the entropy production of the bulk $\dot{s}^{(0)}$ decreases but also the EPR due to the imperfection at $r=0$ does. 

\subsubsection{Spatial profile of the entropy production generated by the impurity}

To obtain explicitly the spatial profile of the entropy production, we consider $\mathcal{G}(r)$. 
Since we could not find an exact expression for it,  we evaluated the integral \eqref{eq:I0p2}
by approximating $\tau^2\tau_I/(\tau+\tau_I) \omega^2(\mathbf{q})\approx \xi^2 q^2$ and extending the 
integration to the whole reciprocal space.
Within this approximation,
the resulting formula reads:
\begin{equation}
\label{eq:I(r)}
\mathcal{G}(r)\approx K_0(r/\xi) \approx \left( \frac{\pi \xi }{2 r}\right)^{1/2} e^{-r/\xi}
\end{equation}
where $K_0(r/\xi)$ is the zeroth-order modified Bessel function of the second kind
\cite{abramowitz1988handbook} and the length $\xi$ is given by Eq.~\eqref{eq:correlationlength_2D}.
The divergence at $r=0$ is the result of the absence of an upper cutoff in the $q$-integral caused 
by the replacement of the Brillouin zone with an infinite integration domain. Since this problem
can be easily fixed by recalling the exact evaluation of $\mathcal{G}(0)$, Eq. \eqref{eq:G0principal}, we have an estimate
of the long distance ($r \gg \xi$) exponential behavior of $\mathcal{G}(r)$.
%

\begin{figure}[!t]
\centering
\includegraphics[width=0.9\linewidth,keepaspectratio]{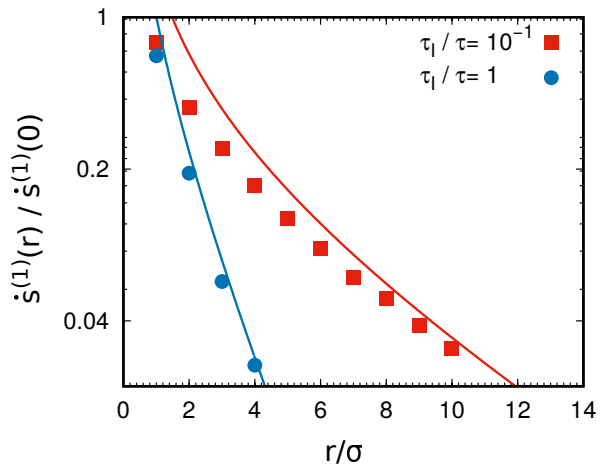}
\caption{
Spatial profile of the EPR, $\dot{s}^{(1)}(r)/\dot{s}^{(1)}(0)$ as a function of the distance from the defect, $r/\sigma$, 
calculated for two different values of the reduced inertial time $\tau_I/ \tau$ as indicated in the plot.
Points are obtained by numerically integrating the dynamics~\eqref{eq:activedynamics} with $|\delta m/m|=0.2$, while the solid colored lines are calculated from the theoretical prediction~\eqref{eq:s1(r)}.
The EPR profiles, $\dot{s}^{(1)}(r)$, 
are estimated from the difference $\dot{s}(r)-\dot{s}^{(0)}$ between the numerical value, $\dot{s}(r)$ and the theoretical bulk value $\dot{s}^{(0)}$. The parameters of the simulations were fixed at $N=10^4$ and $\phi=1.1$.
}\label{fig:Fig3}
\end{figure}

By using the approximation~\eqref{eq:I(r)}, the first order correction to the profile of $\dot{s}(r)$ is given by
\begin{equation}
\begin{aligned}
\label{eq:s1(r)}
\dot{s}^{(1)}(r) &= - \frac{\tau_I}{\tau_I+\tau}\dot{s}_{\text{free}} \left[ K_0(r/\xi) \right]^2 \\
&\approx - \frac{\tau_I}{\tau_I+\tau}\dot{s}_{\text{free}}\left( \frac{\pi \xi }{2 r}\right) e^{-2r/\xi}\, .
\end{aligned}
\end{equation}
It displays an exponential-like decay towards its bulk value $0$, since the EPR of the homogeneous crystal is just $\dot{s}^{(0)}$.
The spatial profile of $\dot{s}^{(1)}(r)$ is shown in Fig.~\ref{fig:Fig3} for two values of the reduced inertial time $\tau_I/\tau$ and reveals a good agreement between the prediction~\eqref{eq:s1(r)} and simulations where $\dot{s}^{(1)}(r)$ was estimated from the difference $(\dot{s}(r)-\dot{s}^{(0)})$, which is exact except for terms order $(\delta m/m)^2$.

The spatial profile of $\dot{s}(r)$ indicates that $\xi/2$ is the typical distance beyond which the entropy production is unchanged by the presence of the impurity: therefore, for large values of $\xi$ an impurity affects the entropy production of the crystal even at long distances from its position.
We recall that $\xi$ increases as $\tau$ for $\tau/\tau_I \ll 1$ (in the underdamped regime), and as $\xi \sim \sqrt{\tau}$ for $\tau/\tau_I \gg 1$ (in the overdamped regime), while in general the value of $\xi$ decreases as the inertia is increased (with the growth of $\tau_I$).
Interestingly, $\xi/2$ coincides with half of the correlation length of the spatial velocity correlations characterizing active crystals.
The correlation between the velocities of two distant particles means a transfer of information that is reflected in the value of the entropy production.

Let us remark that the sign of the first-order correction to $\dot{s}(r)$ depends on the sign of $\delta m$: a defect of the crystal with a mass larger than the one of the remaining particles (heavy impurity) decreases the total entropy production of the system, while an impurity with smaller mass (light impurity) increases the total entropy production.
This phenomenon occurs because a light impurity means more freedom to move for the particles close to the imperfection,
whereas  a heavy impurity reduces the amplitude of their displacements.
Regarding higher order terms in the perturbative expansion, 
it is possible to carry on the calculation using the same method as reported in appendix \ref{appendixsecondorder}
where we have derived a theoretical formula valid for two imperfections up to order $(\delta m/m)^2$.
 However, its application to the present problem is complicated because one would need higher statistics
in the numerical simulations.
This task remains to be carried out in future work.

\section{Conclusions}\label{sec:conclusions}

As shown in the recent literature the EPR is an important theoretical tool to characterize the physics of active and living systems which often involve the presence of many degrees of freedom. However, in the case of nonuniform systems
such a tool becomes much more accurate if instead of employing a global description in terms of the total EPR we use its local version, the space dependent EPR.
In general, this goal is complicated to achieve because of the high dimensionality of the state space. 
In this paper, we have analytically and numerically investigated the space dependent EPR in the case of a active solid with point imperfections 
These lattice defects break the discrete translational symmetry of the crystal and induce a spatial dependence
of the observable quantities. 
This work provides an explicit representation of the spatial variation of the EPR in a very simple model of a nonuniform system. 
Our analytical results for the entropy production in the non-equilibrium steady-state of an active harmonic solid, derived
utilizing a perturbative expansion in powers of the mass of the impurity, agree fairly well  with the numerical data obtained
by computer simulation of the active solid with a soft repulsive shift and cut WCA interparticle potential and ABP dynamics. 
The non-uniform profile of the EPR has been calculated as a function of the distance from the impurity position.
The zeroth-order result of the perturbation theory gives the value of the entropy production of a homogeneous solid~\cite{caprini2022entropons}, whose explicit expression has been reported in the case of a two-dimensional active crystal, while the first-order result accounts for the spatial dependence of the entropy production.
Our study demonstrates that, in regimes of large persistence, an impurity 
affects the properties of the crystal also at a large distance from the impurity. 
Interestingly, the typical length scale that rules the spreading of information, described by the EPR,
is the same length that controls the extent of the spatial velocity correlations in active crystals~\cite{caprini2020hidden, caprini2021spatial}.
Finally, we remark that the lattice imperfection is the cause of the non uniform EPR, but would not cause any velocity correlation profile in the case where only thermal noise would be present. 
This is in agreement with the argument that $\langle{\bf v}_p \cdot {\bf v}_j\rangle=2T \delta_{ij}$
for any equilibrium system.
Future work should extend our approach to binary and polydisperse crystals, to active crystals in three dimensions and to different kind of defects (dislocations, disclinations, grain boundaries)~\cite{chaikin1995principles}.

\section*{Data availability statement}
All data that support the findings of this study are included within the article (and any supplementary
files).

\section*{acknowlegments}

This paper is dedicated to the memory of Luis Felipe Rull (1949-2022).\\

\noindent
LC acknowledges support from the Alexander
Von Humboldt foundation, while HL acknowledges support
by the Deutsche Forschungsgemeinschaft (DFG) through
the SPP 2265 under the grant number LO 418/25-1.

\begin{widetext}

\appendix

\section{Fourier transforms}
\label{appendix:spatial_fouriertransform}

Continuous Fourier transforms in the frequency domain $\omega$ of the generic variable ${\mathbf{O}}_n$ are defined 
as:
\begin{equation}
\hat{\mathbf{O}}_n(\omega)= \int_{-\infty}^\infty dt e^{i\omega t} {\mathbf{O}}_n(t)
\end{equation}
and its inverse is
\begin{equation}
{\mathbf{O}}_n(t)= \int_{\infty}^\infty \frac{d\omega}{2\pi} e^{-i\omega t} \hat{\mathbf{O}}_n(\omega)
\end{equation}
We also employ, in order to decouple the dynamical degrees of freedom, the following discrete spatial Fourier transform of a lattice variable ${\mathbf{O}}_{\mathbf{n}}$
\begin{equation}
 {\mathbf{O}}_{\mathbf{q}}=\frac{1}{ N}\sum_{\mathbf{n}}    {\mathbf{O}}_{\mathbf{n}}  e^{-i \mathbf{q}\cdot \mathbf{r}^0_{\mathbf{n}}   }
\label{fourierrepresentation}
\end{equation}
where $\mathbf{n}$ spans the indices identifying the $N$ particles of the lattice.
Using both the time and lattice Fourier transform
the perfect lattice dynamics ($m_{\mathbf{n}}=m$) takes a particularly simple form since in this representation
each q-mode (wave) becomes independent from the remaining modes and we can write:
\begin{equation}
\label{eq:activedynamicsb}
-m \omega^2 \tilde {\mathbf{u}}_\mathbf{q}= -i\omega \gamma \tilde{\mathbf{u}}_\mathbf{q} -m \omega^2(\mathbf{q}) \tilde {\mathbf{u}}_\mathbf{q} +\sqrt{2 T \gamma}\, \tilde {\boldsymbol{\xi}}_{\mathbf{q}}+\tilde{ \mathbf{f}}^a_{\mathbf{q}}
\end{equation}
where the tilde symbol stands for the  Fourier time and spatial simultaneous transformations
and
\begin{equation}
\omega^2(\mathbf{q})=2\omega_E^2 \Bigl[ 3-\cos(q_x  \bar{x})-2\cos(\frac{1} {2} q_x  \bar{x})\cos (\frac{\sqrt 3} {2} q_y  \bar{x})\Bigr]
\end{equation}
Such a structure function representing the dispersion relation of the triangular lattice is obtained as follows.
One considers the six vectors giving the displacements connecting an arbitrary lattice site to
its  nearest neighbors:
\begin{equation}
 {\bf s}(n_1,n_2)= n_1 {\bf a}_1+n_2 {\bf a}_2
\end{equation}
where
\begin{eqnarray}&&
(n_1,n_2)=\pm( 1,0)\\&&
(n_1,n_2)=\pm(0, 1)\\&&
(n_1,n_2)=\pm( 1,1)
\end{eqnarray}
and ${\bf a}_1,{\bf a}_2$ are the two primitive vectors of the triangular Bravais lattice:
 \begin{eqnarray}&&
{\bf a}_1=\frac{ \bar{x}}{2} {\bf \hat x}-\frac{ \sqrt 3}{2}  \bar{x}{\bf \hat y} \\&&
{\bf a}_2=\frac{ \bar{x}}{2} {\bf \hat x}+\frac{ \sqrt 3}{2}  \bar{x} {\bf \hat y}
\end{eqnarray}
The dispersion relation is obtained by performing the following sum:
\begin{equation}
\omega^2(\mathbf{q})=\omega_E^2 \Bigl(6-\sum_{n_1,n_2} \exp(-i \mathbf{q} \cdot {\bf s}(n_1,n_2))\Bigr)
\end{equation}

\section{Variable transformation}
\label{appendix:change}


In order to perform the integration  over the wavevector ${\bf q}$  in Eq.\eqref{eq:g00} it is convenient, following the literature \cite{guttmann2010lattice}, perform the following change of variables 
and define two new phases $k_1$ and $k_2$ via the transformation:
\begin{eqnarray}&&
k_1=\frac{1} {2} q_x \bar{x}- \frac{\sqrt 3} {2} q_y \bar{x}\\&&
k_2=\frac{1} {2} q_x \bar{x}+\frac{\sqrt 3} {2} q_y \bar{x}.
\end{eqnarray}


With this mapping, the ${\bf q}$-integration domain, that is the hexagonal Brillouin-zone, 
becomes a square domain 
and the resulting integrations
analytically simpler: we first
substitute these relations in eq.~\eqref{eq:omegaq_twoDtriangular} and set  $\mathbf{k}\equiv(k_1,k_2)$:
\begin{equation}
\label{eq:omegaq_twoDtriangularc}
\omega^2(\mathbf{q})\to 
 6\omega_E^2 \Bigl(1-r(\mathbf{k})\Bigr) .
\end{equation}
The last equality defines the new function,
$r(\mathbf{k})=\Bigl(\cos(k_1+k_2) + \cos(k_1)+\cos(k_2 ) \Bigr) $, the so-called structure function of the triangular lattice in two-dimensions.
Thus we rewrite
\begin{equation}
\mathcal{G}(0)= \int_{\Omega} \frac{d{\mathbf{q}}}{\Omega} \frac{1}{1+ \frac{\tau^2\tau_I}{\tau+\tau_I} \omega^2(\mathbf{q})} \,
=   \frac{1}{1+ \frac{ 6 \omega_E^2\tau^2}{1+\tau/\tau_I}} \int_{-\pi}^{\pi}\frac{dk_1}{2\pi} \int_{-\pi}^{\pi} \frac{dk_2}{2\pi} \frac{1}{1- \frac{1+\tau/\tau_I}{ 6 \omega_E^2 \tau^2}            r(\mathbf{k}) }
\end{equation}
The result of the last integration can be found in Ref.~\cite{guttmann2010lattice} and reads:
\begin{equation}
\label{eq:G0}
\mathcal{G}(0)=\frac{1}{1+\xi^2} \frac{6}{\pi z(\xi) \sqrt{c(\xi)}} K[k(\xi)]
\end{equation}
where $K(k)$ is the complete elliptic integral of the first kind and $z(\xi)$, $c(\xi)$ and $k(\xi)$ are explicit functions of the parameters of the model and are given by:
\begin{flalign}
& z(\xi)=\frac{1}{1+\frac{\bar{x}^2}{4\xi^2}}\\
&c(\xi)=\frac{9}{z(\xi)^2}-3+2\sqrt{3+\frac{6}{z(\xi)}}\\
&k(\xi)=2\frac{\left(3+ \frac{6}{z(\xi)}\right)^{1/4}}{c(\xi)^{1/2}} \,.
\end{flalign}

$$
\xi^2= \frac{3}{2} \bar{x}^2 \frac{\tau^2}{1+\tau/\tau_I }\omega_E^2\, .
$$

\section{Lattice Green's function and perturbative solution}
\label{appendix_pntegrals_omega}
Let us multiply
Eq.~\eqref{eq:Fourierspace_dynamics} at the right by the inverse operator
$L^{-1}=G$ such that $G_{pn}L_{nk}=\delta_{pk}$ and get:
\begin{equation}
\label{eq:Fourierspace_dynamicsinverse}
\hat{\mathbf{u}}_p(\omega) =G_{pn}(\omega)\Bigl( \hat{\mathbf{f}}^a_n + \sqrt{2 \gamma T} \hat{\boldsymbol{\xi}}_n\Bigr)+
 \lambda m \omega^2 G_{p0} \hat{\mathbf{u}}_0  
\end{equation}
where $\lambda$ is a small perturbative parameter.
This equation is solved by iteration using  $\lambda$ as a smallness parameter:
\begin{equation}
\hat{\mathbf{u}}_p =\Bigl(  G_{pm}(\omega)+\lambda  G_{pn}(\omega) b_n(\omega)    G_{nm}(\omega) + \lambda ^2 G_{pn}(\omega) b_n(\omega) G_{nk}(\omega)  b_k(\omega)    G_{km}(\omega)+\dots\Bigr) 
\Bigl( \hat{\mathbf{f}}^a_m + \sqrt{2 \gamma T} \hat{\boldsymbol{\xi}}_m\Bigr)
\end{equation}
where $b_n=m\omega^2 \delta_{n0}$ for a single impurity sitting at site $0$.
The explicit representation of the matrix element $G_{pn}(\omega)$ is obtained by first solving the
homogeneous eigenvalue problem:
\begin{equation}
  L_{pn}(\omega)  \phi_{\mathbf{n}}(\mathbf{q})= \lambda(\mathbf{q},\omega)  \phi_{\mathbf{p}}(\mathbf{q})
 \label{eigenvalueequation}
 \end{equation}
with eigenvalues
 \begin{equation}
 \lambda(\mathbf{q},\omega)=-m\omega^2 +i\omega\gamma+ m\omega^2(\mathbf{q})
 \end{equation}
and eigenvectors
$$
 \phi_{\mathbf{n}}(\mathbf{q})= \frac{1}{\sqrt N}e^{-i \mathbf{q}\cdot \mathbf{r}^0_{\mathbf{n}}   }
 $$
 that depend on the wavevector $\mathbf{q}$.
The resolvent or Greens's function has the following spectral representation:
\begin{equation}
G_{pn}(\omega)=\sum_\mathbf{q} \frac{1}{ -m\omega^2 +i\omega \gamma+ m\omega^2(\mathbf{q})}    \phi^*_{\mathbf{p}}(\mathbf{q})  \phi_{\mathbf{n}}(\mathbf{q})
\end{equation}

\section{Active Ornstein-Uhlenbeck approximation and derivation of Eq.~\eqref{eq:fafa_corr}}\label{appendix:AOUP}

Here, we discuss the argument justifying the replacement of the ABP dynamics with the AOUP dynamics. 
The motivation is twofold: i) the theoretical manipulation of the AOUP equations of evolution is simpler than
the ABP equations used in the numerical work, ii) there is a correspondence between the two models based on the property that the respective autocorrelation functions of the active force have the same form.
To prove the second statement, consider
 that the ABP dynamics of $\mathbf{f}^a_n$, described by Eqs.~\eqref{eq:self-propulsiondef} and~\eqref{eq:dynamica_theta},
 can be expressed in Cartesian coordinates as~\cite{marconi2021hydrodynamics,caprini2022parental}: 
\begin{equation}
\dot{\mathbf{f}}^a_n = -  D_r \mathbf{f}^a_n +  \gamma v_0\sqrt{2 D_r} \boldsymbol{\eta}_n\times \mathbf{f}^a_n \,,
\end{equation}
where we adopt the Ito convention to interpret the second term in the r.h.s..
The noise vector $\boldsymbol{\eta}_n=(0, 0, \eta_n)$ is normal to the plane of motion $(x, y)$  $\eta_n$  isa white noise with zero average and unit variance. It is easy to show (see Ref.~\cite{farage2015effective}) that
the autocorrelation function of $\mathbf{f}^a_p$  is an exponential of the form:
\begin{equation}
\
\langle \mathbf{f}^a_n(t) \cdot \mathbf{f}^a_m(0) \rangle = \delta_{nm} v_0^2\gamma^2 e^{-|t| /\tau}\, 
\end{equation}
where $\tau=1/D_r$. As anticipated, in theoretical work, it is convenient to modify
the dynamics of $f^a$ while preserving the form of its autocorrelation function.
This goal is achieved
by replacing the ABP noise term, $\sqrt{2 D_r} \boldsymbol{\eta}_n\times \mathbf{f}^a_n$, by a two-dimensional white noise
vector of white noises, $\boldsymbol{\zeta}_n$ such that $\langle \boldsymbol{\zeta}_n(t) \boldsymbol{\zeta}_m(0) \rangle=\delta_{nm} \delta(t)$.
This replacement corresponds to approximate the dynamics of $\mathbf{f}^a_p$ by the following two-dimensional Ornstein-Uhlenbeck process
\begin{equation}
\label{eq:AOUPdynamics_app}
\dot{\mathbf{f}}^a_n = - D_r \mathbf{f}^a_n + \gamma v_0 \sqrt{2 D_r} \boldsymbol{\zeta}_n.
\end{equation}
It is easy to verify that this new equation for the active force yields the same autocorrelation function as the ABP model.
Equation~\eqref{eq:AOUPdynamics_app} together with Eq.~\eqref{eq:activedynamics} corresponds to the active Ornstein-Uhlenbeck particles (AOUP) model.

The dynamics~\eqref{eq:AOUPdynamics_app} in the Fourier Space (in the frequency $\omega$ domain), can be obtained by applying the Fourier transform and is given by
\begin{subequations}
\label{eq:app_qdyn}
\begin{align}
i\omega \tau \hat{ \mathbf{f}}^a_n(\omega) = -\hat{ \mathbf{f}}^a_n(\omega)  + v_0\gamma\sqrt{2\tau} \hat{\boldsymbol{\zeta}}_n(\omega)  \,,
\end{align}
\end{subequations}
and allows to find the explicit solution for $\hat{ \mathbf{f}}^a_n(\omega)$ as
\begin{equation}
\hat{ \mathbf{f}}^a_n(\omega) = \frac{v_0 \gamma \sqrt{2\tau} }{1+i \omega \tau}  \hat{\boldsymbol{\zeta}}_n(\omega) \,.
\end{equation}
By multiplying by $\hat{ \mathbf{f}}^a_m(-\omega)$, taking the average over the realizations of the noise, dividing by $t$, and applying the limit $t\to\infty$, we obtain Eq.~\eqref{eq:fafa_corr}.

\section{Integrals over the frequency domain}
\label{Integralsfrequency}
This appendix contains some details about the derivation of the formula for  first order correction to the EPR. 
We first compute the following integral over frequencies which appears in
Eq.\eqref{eq:zerothspectralentropy}
\begin{equation}
I_2(\mathbf{q}) = \int_{-\infty}^\infty  \frac{d\omega}{2\pi} \frac{1}{1+\omega^2\tau^2}\, \frac{\omega^2 }{m^2 (\omega^2(\mathbf{q})- \omega^2)^2 + \omega^2 \gamma^2 }=  \frac{1}{2\gamma m}  \frac{1}{ \Bigl[1+\gamma\tau/m+\tau^2\omega^2(\mathbf{q})\Bigr]}\\
\label{eq:i2integral}
\end{equation}
\subsection{Zeroth order EPR}
Using the result \eqref{eq:i2integral} we evaluate the zeroth order entropy production:
$$
T\dot{s}^{(0)} = \frac{2}{N} \sum_{q}  \int_{-\infty}^\infty \frac{d\omega}{2\pi} \frac{\tau\gamma}{1+\omega^2\tau^2} \frac{\omega^2 v_0^2 \gamma^2}{m^2 (\omega^2(\mathbf{q})- \omega^2)^2 + \omega^2 \gamma^2 }
=\frac{ v_0^2 \gamma }{ m}\tau\gamma\,  \frac{1}{N} \sum_{q} \frac{1}{ \Bigl[1+\gamma\tau/m+\tau^2\omega^2(\mathbf{q})\Bigr]}
$$

$$
T\dot{s}^{(0)} 
= v_0^2 \tau\gamma\, \frac{1}{ \tau+\tau_I} \frac{1}{N} \sum_{q}  \frac{1}{ \Bigl[1+\frac{\tau^2\tau_I\omega^2(\mathbf{q})}{\tau+\tau_I}\Bigr]}
$$
By using (see Eq.\eqref{eq:I0p}) 
\begin{equation}
\mathcal{G}_{0p} =\frac{1}{N} \sum_q \frac{e^{-i \mathbf{q}\cdot \mathbf{r}_p^0}}{1+ \frac{\tau^2\tau_I}{\tau+\tau_I} \omega^2(\mathbf{q})}
\to \int_{\Omega} \frac{d{\mathbf{q}}}{\Omega} e^{-i \mathbf{q}\cdot \mathbf{r}_p^0}\,
 \frac{1}{1+ \frac{\tau^2\tau_I}{\tau+\tau_I} \omega^2(\mathbf{q})}
 \nonumber
\end{equation}
We arrive at
$$
T\dot{s}^{(0)} 
= v_0^2 \tau\gamma\, \frac{1}{\tau+\tau_I} \mathcal{G}_{00}
$$
\subsection{First order EPR}
According to the second equality in Eq.\eqref{eq:first_orderEP_2}
We now consider 
the first order correction.  We need to perform the following frequency integral
\begin{eqnarray}\\&&
\label{eq:integralfirstorder}
T\dot{s}^{(1)}_p= -2 mv_0^2 \gamma^2 \tau
\int_{-\infty}^\infty \frac{d  \omega}{2\pi}\frac{\omega^3 }{1+\omega^2\tau^2} \text{Im} [G_{p0}(\omega) G_{0p}(\omega)] = \nonumber\\&&
 -\frac{2 mv_0^2 \gamma^2 \tau}{N^2}\sum_{\mathbf{q},\mathbf{q}'} e^{i (\mathbf{q}- \mathbf{q'})\cdot \mathbf{r}^0_{\mathbf{p}}   }
  \int_{-\infty}^\infty \frac{d  \omega}{2\pi}\frac{\omega^3 }{1+\omega^2\tau^2} \text{Im} [ \frac{1}{ m\omega^2 -i\omega \gamma- m\omega^2(\mathbf{q})} 
   \frac{1}{ m\omega^2 -i\omega \gamma- m\omega_\mathbf{q'}^2}  ]  
\end{eqnarray}
We use the following identity
$$
\Bigl( \frac{1} {\omega^2-\omega^2_{q}-i\omega \gamma/m}  \Bigr)\Bigl(   \frac{1}{\omega^2-\omega^2_{q'}-i\omega \gamma/m}    \Bigr) 
=\frac{1}{\omega^2_q-\omega^2_{q'}}\Bigl( \frac{1} {\omega^2-\omega^2_{q}-i\omega \gamma/m}  -  \frac{1}{\omega^2-\omega^2_{q'}-i\omega \gamma/m}    \Bigr) 
$$
After multiplying the previous expression by $\frac{\omega^3 }{1+\omega^2\tau^2} $ and taking its imaginary part,
we arrive at the following form of the frequency integral appearing in Eq.\eqref{eq:integralfirstorder}
which is evaluated by residue theorem method:
\begin{equation}
\begin{aligned}
& \frac{ \gamma/m  }{\omega^2(\mathbf{q})-\omega_\mathbf{q'}^2} \int_{-\infty}^\infty \frac{d  \omega}{2\pi}
 \frac{   \omega^4   }{1+\omega^2 \tau^2} \Bigl(
 \frac{1} {(\omega^2-\omega^2(\mathbf{q}))^2+\omega^2 \gamma^2/m^2} -
 \frac{1} {(\omega^2-\omega_\mathbf{q'}^2)^2+\omega^2 \gamma^2/m^2} 
  \Bigr)\\
&\qquad\qquad\qquad\qquad\qquad\qquad\qquad\qquad\qquad\qquad=\frac{1}{2 }\frac{ 1}
 {( 1+\gamma\tau/m+\omega^2(\mathbf{q})\tau^2)( 1+\gamma\tau/m+\omega_\mathbf{q'}^2\tau^2 )}
\end{aligned}
\end{equation}
Finally, inserting this result in Eq.\eqref{eq:integralfirstorder} and performing separately the two 
independent integrations over $\mathbf{q'}$ and  $\mathbf{q}$  we find:
\begin{eqnarray}\\&&
 \int_{-\infty}^\infty \frac{d  \omega}{2\pi}\frac{\omega^3 }{1+\omega^2\tau^2} \text{Im} [G_{p0}(\omega) G_{0p}(\omega)] = 
\nonumber\\&&
= 
\frac{1}{m^2} \frac{1}{2N^2}\sum_{\mathbf{q},\mathbf{q}'} 
  \frac{ e^{i (\mathbf{q}- \mathbf{q'})\cdot \mathbf{r}^0_{\mathbf{p}}   }}{( 1+\tau/\tau_I)^2}  \frac{1}{1+ \frac{\tau^2\tau_I}{\tau+\tau_I} \omega^2(\mathbf{q})}
  \frac{1}{1+ \frac{\tau^2\tau_I}{\tau+\tau_I} \omega^2(\mathbf{q'})}
 =\frac{1}{2m^2}   \frac{ 1}{( 1+\tau/\tau_I)^2} \mathcal{G}_{0p} \mathcal{G}_{p0} 
\end{eqnarray}
where we used Eq.\eqref{eq:I0p}.
Finally, we write:
\begin{equation}
\label{eq:first_orderEP_3}
\begin{aligned}
Ts_p^{(1)}(\omega) = & - 2 m  v_0^2 \gamma^2 \tau\int_{-\infty}^\infty \frac{d  \omega}{2\pi}  \omega^3 \text{Im} [G_{p0}(\omega) G_{0p}(\omega)] \frac{1}{1+\omega^2\tau^2} \, 
= - \frac{v_0^2 \tau \gamma \tau_I}{(\tau + \tau_I)^2}
\mathcal{G}_{0p} \mathcal{G}_{p0} 
\end{aligned}
\end{equation}


\subsection{Calculation of EPR up to Second order}
\label{appendixsecondorder}

Here we consider two impurities at positions $a$ and $b$ and having mass defects $\delta m_a$ and $\delta m_b$, respectively.
To study the EPR due to two defects we need to carry on a second order calculation in the perturbation parameters $\delta m$. Thus we write the expression of the average of the product between the active force and the velocity of a single particle 
as a function of its lattice position, $p$:
\begin{eqnarray}&&
\label{eq:expansionperturbative}
\langle \hat f^a_p(-\omega)  \hat V_p(\omega)\rangle =i\omega 2 v_0^2 \gamma^2  \frac{\tau}{1+\omega^2 \tau^2} \Bigl( \hat  { G}_{pp}(\omega)+\lambda \bigl[ \hat b_a(\omega) {  \hat G}_{pa}(\omega) { \hat  G}_{ap}(\omega) +\hat b_b(\omega)  { \hat  G}_{pb}(\omega) { \hat  G}_{bp}(\omega) \bigr] + 
\nonumber\\&&
\lambda ^2 \bigl[ \hat b_a^2(\omega) { \hat  G}_{pa}(\omega) { \hat  G}_{aa}(\omega){ \hat  G}_{ap}(\omega) +\hat b_b^2(\omega){ \hat  G}_{pb}(\omega)  { \hat  G}_{bb}(\omega) { \hat  G}_{bp}(\omega)  + \hat b_a(\omega) \hat b_b(\omega)  { \hat  G}_{pa}(\omega) { \hat  G}_{ab}(\omega) { \hat  G}_{bp}(\omega) \nonumber \\&&+
 \hat b_a(\omega) \hat b_b(\omega) {  \hat G}_{pb}(\omega)  { \hat  G}_{ba}(\omega) {  \hat G}_{ap}(\omega)
 \bigr]  \Bigr) 
\end{eqnarray}
where
$
 \hat b_a(\omega)=\delta m_a \,\omega^2
$
and we truncate the expansion at the second order in $\lambda$. 
To obtain the second order correction to the EPR, we add 
its complex conjugate (c.c) and divide by a factor $2$ and the resulting formula must be integrated over $\omega$.

Since we have already computed the expansion up to the first order, we here write only the second order correction 
contained in the long expression \eqref{eq:expansionperturbative}. 
In the following, for conciseness, we set $\omega_n^2=\omega^2(q_n)$.

Using the Fourier representation of the lattice Green's functions 
\eqref{eq:greenlattice}
the second order correction involves the following type of integrals over frequencies and sums over wavevectors:
\begin{eqnarray}&&
\int_{-\infty}^\infty \frac{d\omega}{2\pi}
\frac{i\tau \omega^5}{1+\omega^2 \tau^2} 
  { \hat  G}_{p0}(\omega) { \hat  G}_{00}(\omega){ \hat  G}_{0p}(\omega)+c.c
   \nonumber\\&&
  =
-\Bigl(  \frac{1}{N}\frac{1}{m}\Bigr)^3 \sum_{q_1} \sum_{q_2} \sum_{q_3}
\int_{-\infty}^\infty \frac{d\omega}{2\pi}
\frac{i\omega^5\tau }{1+\omega^2 \tau^2} 
(\frac{e^{-i \mathbf{q_1}\cdot \mathbf{r}_p^0} }{\omega^2-i\gamma \omega/m-\omega^2_1})(\frac{1}{\omega^2-i\gamma \omega/m-\omega^2_2})(\frac{e^{i \mathbf{q_3}\cdot \mathbf{r}_p^0}}{\omega^2-i\gamma \omega/m-\omega^2_3})+c.c.\nonumber\\&&
\label{eq:secondorder}
\end{eqnarray}
Using the residue theorem to perform the  above $\omega$-integral
we arrive at:
\begin{equation}
\begin{aligned}
\int_{-\infty}^\infty \frac{d\omega}{2\pi}
\langle \hat f^a_p(-\omega)  \hat V_p(\omega)\rangle &=
 v_0^2\gamma^2 \tau (\delta m)^2 \Bigl(  \frac{1}{N}\frac{1}{m}\Bigr)^3 \sum_{q_1} \sum_{q_2} \sum_{q_3}  e^{-i \mathbf{q_1}\cdot \mathbf{r}_p^0} e^{-i \mathbf{q_3}\cdot \mathbf{r}_p^0}\times\\
&
 \times\Bigl\{   
  \frac{1}{(1+\tau/\tau_I+\omega_1^2\tau^2) \, (1+\tau/\tau_I+\omega_2^2\tau^2) \, (1+\tau/\tau_I+\omega_3^2\tau^2)
 } \Bigr\} 
\end{aligned}
\end{equation}
Gathering all together, we 
find the following nice formula for the EPR:
\begin{eqnarray}&&
\sum_{n} \dot{s}_p = \frac{1}{2T}  \sum_{p} \Bigl( \langle V_p(\omega) f_p^a(-\omega) \rangle+c.c\Bigr)= \frac{1}{T} \frac{ v_0^2\gamma^2}{m} \tau \sum_{n} \Bigl\{\frac{1}{1 +\tau/\tau_I   }   \, {\mathcal{G}}_{pp}- 
( \frac{1}{1 +\tau/\tau_I   })^2   \Bigl(\frac{\delta m_a}{ m}{\mathcal{G}}_{ap} {\mathcal{G}}_{pa}+\frac{\delta m_b}{ m}{\mathcal{G}}_{bp} {\mathcal{G}}_{pb} \Bigr) \nonumber\\&&
+  (\frac{1}{1 +\tau/\tau_I   } )^3
 \Bigl( \frac{\delta m_a^2}{m^2}{\mathcal{G}}_{pa}  {\mathcal{G}}_{aa}  {\mathcal{G}}_{ap}+
 \frac{\delta m_b^2}{m^2}{\mathcal{G}}_{pb}  {\mathcal{G}}_{bb}  {\mathcal{G}}_{bp}
+2\frac{\delta m_a \delta m_b}{m^2}{\mathcal{G}}_{pa}  {\mathcal{G}}_{ab}  {\mathcal{G}}_{bp} \Bigr)\Bigr\}+\dots
\end{eqnarray}

The propagators ${\mathcal{G}}_{aa} $ and ${\mathcal{G}}_{bb}$  are equal but the prefactors depend on the signs
and amplitudes of the mass ratios.
 Switching to continuous notation $\mathcal{G}_{ab}\to{\mathcal{G}}(r_{ab})$. It decays exponentially with the separation
 $r_{ab}$ between two lattice points and a characteristic length $\xi$.
Perhaps, the most interesting term is the last because describes the local value of the EPR (i.e. at ${\bf r}_p$) due to the combined effect of two imperfections sitting at ${\bf r}_a$ and ${\bf r}_b$, respectively.

\end{widetext}

\bibliographystyle{apsrev4-1}

\bibliography{EP.bib}

\end{document}